\documentclass[longauth,traditabstract,letter]{aa}
\pdfoutput=1

\usepackage{graphicx}
\usepackage{txfonts}
\usepackage{natbib}
\bibpunct{(}{)}{;}{a}{}{,}
\usepackage{hyperref}
\usepackage{ucs}
\usepackage[utf8x]{inputenc}
\usepackage[T1]{fontenc}
\newcommand{\herschel}{{\it Herschel}}
\newcommand\hiioii{\ifmmode{\rm H_2O_2}\else H$_2$O$_2$\fi}
\newcommand\coii{\ifmmode{\rm CO_2}\else CO$_2$\fi}
\newcommand\oii{\ifmmode{\rm O_2}\else O$_2$\fi}
\newcommand\xiiico{\ifmmode{\rm ^{13}CO}\else $^{13}$CO\fi}
\newcommand\cxviiio{\ifmmode{\rm C^{18}O}\else C$^{18}$O\fi}
\newcommand\xiicxviiio{\ifmmode{\rm ^{12}C^{18}O}\else $^{12}$C$^{18}$O\fi}
\newcommand\kms{\ifmmode{\rm km\thinspace s^{-1}}\else km\thinspace s$^{-1}$\fi}
\graphicspath{{./figures/}{/home/miguel/HssO/Mars/CO-paper/figures/}}

\begin{document}

\titlerunning{\herschel{}/HIFI observations of HCl, \hiioii{} and \oii{} in the
Martian atmosphere}
\title{\herschel{}/HIFI observations of Mars: first detection of
\oii{} at submillimetre wavelengths and upper limits on HCl and \hiioii{}%
\thanks{\herschel{} is an ESA space observatory with science instruments
provided by European-led Principal Investigator consortia and with
important participation from NASA.}}

\author{
  P.~Hartogh\inst{\ref{inst1}}
  \and C.~Jarchow\inst{\ref{inst1}}
  \and E.~Lellouch\inst{\ref{inst2}}
  \and M.~de~Val-Borro\inst{\ref{inst1}}
  \and M.~Rengel\inst{\ref{inst1}}
  \and R.~Moreno\inst{\ref{inst2}}
  \and A.~S.~Medvedev\inst{\ref{inst1}}
  \and H.~Sagawa \inst{\ref{inst1},\ref{inst28}}
  \and B.~M.~Swinyard\inst{\ref{inst9}}
  \and T.~Cavali\'e\inst{\ref{inst1}}
  \and D.~C.~Lis\inst{\ref{inst3}}
  \and M.~I.~B\l{}\k{e}cka\inst{\ref{inst4}}
  \and M.~Banaszkiewicz\inst{\ref{inst4}}
  \and D.~Bockel\'ee-Morvan\inst{\ref{inst2}}
  \and J.~Crovisier\inst{\ref{inst2}}
  \and T.~Encrenaz\inst{\ref{inst2}}
  \and M.~K\"uppers\inst{\ref{inst7}}
  \and L.-M.~Lara\inst{\ref{inst8}}
  \and S.~Szutowicz\inst{\ref{inst4}}
  \and B.~Vandenbussche\inst{\ref{inst10}}
  \and F.~Bensch\inst{\ref{inst5}}
  \and E.~A.~Bergin\inst{\ref{inst11}}
  \and F.~Billebaud\inst{\ref{inst25}}
  \and N.~Biver\inst{\ref{inst2}}
  \and G.~A.~Blake\inst{\ref{inst3}}
  \and J.~A.~D.~L.~Blommaert\inst{\ref{inst10}}
  \and J.~Cernicharo\inst{\ref{inst12}}
  \and L.~Decin\inst{\ref{inst10},\ref{inst23}}
  \and P.~Encrenaz\inst{\ref{inst13}}
  \and H.~Feuchtgruber\inst{\ref{inst21}}
  \and T.~Fulton\inst{\ref{inst20}}
  \and T.~de~Graauw\inst{\ref{inst18},\ref{inst24},\ref{inst14}}
  \and E.~Jehin\inst{\ref{inst6}}
  \and M.~Kidger\inst{\ref{inst15}}
  \and R.~Lorente\inst{\ref{inst15}}
  \and D.~A.~Naylor\inst{\ref{inst16}}
  \and G.~Portyankina\inst{\ref{inst19}}
  \and M.~S\'anchez-Portal\inst{\ref{inst15}}
  \and R.~Schieder\inst{\ref{inst17}}
  \and S.~Sidher\inst{\ref{inst9}}
  \and N.~Thomas\inst{\ref{inst19}}
  \and E.~Verdugo\inst{\ref{inst15}}
  \and C.~Waelkens\inst{\ref{inst10}}
  \and N.~Whyborn\inst{\ref{inst14}}
  \and D.~Teyssier\inst{\ref{inst15}}
  \and F.~Helmich\inst{\ref{inst18}}
  \and P.~Roelfsema\inst{\ref{inst18}}
  \and J.~Stutzki\inst{\ref{inst17}}
  \and H.~G.~LeDuc\inst{\ref{inst26}}
  \and J.~A.~Stern\inst{\ref{inst26}}
  }

\institute{
  Max-Planck-Institut f\"ur Sonnensystemforschung, 37191
  Katlenburg-Lindau, Germany\label{inst1}
  \and LESIA, Observatoire de Paris, 5 place Jules Janssen, 92195
    Meudon, France\label{inst2}
  \and Environmental Sensing \& Network Group, NICT,
  4-2-1 Nukui-kita, Koganei, Tokyo 184-8795, Japan\label{inst28}
  \and STFC Rutherford Appleton Laboratory, Harwell Innovation
    Campus, Didcot, OX11 0QX, UK\label{inst9}
  \and California Institute of Technology, Pasadena, CA 91125, USA\label{inst3}
  \and Space Research Centre, Polish Academy of Sciences, Warsaw,
    Poland\label{inst4}
  \and Rosetta Science Operations Centre, European Space Astronomy
    Centre, European Space Agency, Spain\label{inst7}
  \and Instituto de Astrof\'isica de Andaluc\'ia (CSIC), Spain\label{inst8}
  \and Instituut voor Sterrenkunde, Katholieke Universiteit Leuven,
    Belgium\label{inst10}
  \and DLR, German Aerospace Centre, Bonn-Oberkassel, Germany\label{inst5}
  \and Astronomy Department, University of Michigan, USA\label{inst11}
  \and Universit\'e de Bordeaux, Laboratoire d'Astrophysique de
    Bordeaux, France\label{inst25}
  \and Laboratory of Molecular Astrophysics, CAB-CSIC, INTA, Spain\label{inst12}
  \and Sterrenkundig Instituut Anton Pannekoek, University of Amsterdam,
    Science Park 904, 1098 Amsterdam, The Netherlands\label{inst23}
  \and LERMA, Observatoire de Paris, France\label{inst13}
  \and Max-Planck-Institut f\"ur extraterrestrische Physik, 
    Giessenbachstra\ss e, 85748 Garching, Germany\label{inst21}
  \and Bluesky Spectroscopy, Lethbridge, Canada\label{inst20}
  \and SRON Netherlands Institute for Space Research, Landleven 12, 9747
    AD, Groningen, The Netherlands\label{inst18}
  \and Leiden Observatory, University of Leiden, The Netherlands\label{inst24}
  \and Atacama Large Millimeter/Submillimeter Array, Joint ALMA Office,
    Santiago, Chile\label{inst14}
  \and Institute d'Astrophysique et de Geophysique, Universit\'e de Li\`ege,
    Belgium\label{inst6}
  \and \herschel{} Science Centre, European Space Astronomy
    Centre, ESA, P.O.  Box 78, 28691 Villanueva de la Ca\~nada, Madrid Spain\label{inst15}
  \and Department of Physics and Astronomy, University of Lethbridge,
    Canada\label{inst16}
  \and Physikalisches Institut, University of Bern, Switzerland\label{inst19}
  \and KOSMA, I. Physik. Institut, Universität zu K\"oln, Z\"ulpicher Str.
    77, D 50937 K\"oln, Germany\label{inst17}
  \and Jet Propulsion Laboratory, Caltech, Pasadena, CA 91109,
    USA\label{inst26}
  }

\date{Received May 31, 2010; accepted}

\abstract{We report on an initial analysis of  \herschel{}/HIFI
observations of hydrogen chloride (HCl), hydrogen peroxide (\hiioii{}),
and molecular oxygen (\oii{}) in the Martian atmosphere performed on 13
and 16 April 2010 ($L_s \sim 77\degr$). We derived a constant volume
mixing ratio of $1400 \pm 120$ ppm for \oii{} and determined upper
limits of 200 ppt for HCl and 2 ppb for \hiioii{}. Radiative transfer
model calculations indicate that the vertical profile of \oii{} may not
be constant. Photochemical models determine the lowest values of \hiioii{}
to be around $L_s \sim 75\degr$ but overestimate the volume mixing ratio
compared to our measurements.}


\maketitle

\section{Introduction}

Hydrogen chloride (HCl) is a reservoir of chlorine species and plays an
important role in the atmospheric chemistry of Venus and Earth. Its
detection by ground-based  infrared spectroscopy
\citep{2008P&SS...56.1424I} and space borne UV stellar/solar
occultation observations by SPICAV/SOIR on Venus Express
\citep{2009DPS....41.6002B} provide mid atmospheric mixing ratios
between 0.1 and 1 ppm in the Venusian atmosphere. 
Submillimetre wave observations of HCl in the Earth atmosphere have long been
performed from an airplane
\citep{1994GeoRL..21.1267C,1995JGR...10020957W}. The derived relative
abundances are  $\sim 2$ orders of magnitude smaller than in Venus
($\sim$ 1--3 ppb). In the Martian atmosphere HCl has not been found yet.
Its detection would be an indication of present volcanic activity on
Mars \citep{2003JGRE..108.5026W,2004Icar..170..424E}.
\citet{1997JGR...102.6525K} presented a stringent upper limit of 2 ppb
from high-resolution ground-based observations of Mars.

The situation is somewhat different for hydrogen peroxide (\hiioii{}).
It was detected for the first time in 2003 by
\citet{2004Icar..168..116C} and \citet{2004Icar..170..424E} in the
Martian atmosphere. The observed abundance varied between 20 and 40
ppb, consistent with photochemical model calculations
\citep[e.g.;][]{1993Icar..101..313K,1994JGR....9913133A,1994Icar..111..124N}
for the northern fall season ($L_s = 206\degr$).    \hiioii{} may also
be produced by electrostatic discharge reactions during  dust storms, in
dust devils, or during normal saltation \citep{2006AsBio...6..439A}.
Near the surface, the concentration could exceed 200 times that
produced by photochemistry alone,  enough for  condensation and
precipitation of \hiioii{} to occur. In its solid phase on the surface, it
may be responsible for scavenging organic material from Mars and/or
present a sink of methane such that a larger source is required
to maintain its steady-state abundance
\citep[e.g.][]{2009Sci...323.1041M}.

Oxygen was claimed to be detected  for the first time in the Martian atmosphere
(in addition to water) by \citet{1909Sci....30..678V}. It took
almost 60 years until \citet{1968ApJ...153..963B} tentatively confirmed
the detection of \oii{}  in the oxygen A band (around 763 nm) with a
mixing ratio of 2600 ppm or less. They claimed that the CO/\oii{} ratio
was two, consistent with the assumption that both gases were produced by
the decomposition of \coii{}.  
By performing observations of the same wavelength range,
\citet{1972Natur.238..447B} and \citet{1972Sci...177..988C} found 
only 1300 ppm of \oii{}.  
Since \citet{1969ApJ...157L.187K} had in the meanwhile reported  a reliable
measurement of 800 ppm of CO, they concluded that there was an additional
source of \oii{} namely most likely water. 
Molecular oxygen is a non-condensable species in the
Martian atmosphere. The pressure of the Martian atmosphere oscillates
annually by about a third due to the condensation and sublimation of
\coii{}, i.e. this variation should also appear in the \oii{} volume mixing
ratio.  \citet{2004otp..work.3009E} reanalyzed the Viking
lander data and found variations from 2500 to 3300 ppm. They point out
that the 1300 ppm published by \citet{1977JGR....82.4635O} are not based
on  Viking measurements, but on the ground-based data cited above and
claim that the amount of 3000 ppm is high enough to directly extract
oxygen for use as a propellant for sample or crew return as well as for
the breathing of astronauts \citep{2001England}.

\begin{table*}
  \caption{HIFI observations of HCl, \hiioii{} and \oii{} in Mars.
  }
  \label{table:1} \centering
  \begin{tabular}{c c c c c c c c c}
    \hline\hline OD & Obs.\ ID & Integration time & UT start date & 
    Molecule & Transition & Sideband & Frequency & Beam size\\
    & & [s] & & & & & [GHz] & [$\arcsec$] \\
    \hline
    334 & 1342194690 & 9289 &  2010-04-13 06:39:28  &  O$_2$  &  5,4 $\rightarrow$ 3,4  &  LSB & 773.840 & 27.4 \\
& & &   &  C$^{17}$O &  $7 \rightarrow 6$  &  USB & 786.281 & 27.0 \\
\hline
334 & 1342194689 & 2297 &  2010-04-13 05:59:40  &  O$_2$  &  5,4 $\rightarrow$ 3,4  &  USB & 773.840 & 27.4 \\
\hline
337 & 1342194756 & 2505 &  2010-04-16 14:53:08  &  \hiioii{} &  $ 5 \rightarrow 4$  &  USB & 1847.123 & 11.5 \\
& & &   &  CO  &  $16 \rightarrow 15$  &  LSB & 1841.346 & 11.5 \\
\hline
337 & 1342194755 & 3746 &  2010-04-16 13:48:47  &  HCl  &  3 4 $\rightarrow$ 2 4  &  USB & 1876.211 & 11.3 \\
& & &   &   &  3,3 $\rightarrow$ 2,4  &   & 1876.218 & 11.3 \\
& & &   &   &  3,2 $\rightarrow$ 2,1  &   & 1876.223 & 11.3 \\
& & &   &   &  3,3 $\rightarrow$ 2,2  &   & 1876.223 & 11.3 \\
& & &   &   &  3,4 $\rightarrow$ 2,3  &   & 1876.227 & 11.3 \\
& & &   &   &  3,5 $\rightarrow$ 2,4  &   & 1876.227 & 11.3 \\
& & &   &   &  3,3 $\rightarrow$ 2,3  &   & 1876.235 & 11.3 \\
& & &   &   &  3,2 $\rightarrow$ 2,2  &   & 1876.240 & 11.3 \\
& & &   &   &  3,2 $\rightarrow$ 2,3  &   & 1876.252 & 11.3 \\

    \hline
  \end{tabular}
  \label{table1}
\end{table*}

The observations of the HCl, \hiioii{}, and \oii{} in the Martian
atmosphere are part of the \herschel{} key programme ``Water and related
chemistry in the solar system'' \citep{2009P&SS...57.1596H}.  This paper
describes the observations and data analysis and provides the volume
mixing ratios of the gases and their upper limits. 

\section{\herschel{}/HIFI observations}

The set of HIFI observations was carried out between 11 and 16 April
2010 corresponding to $L_s = 75.8\degr$ to 78°, including spectral line
surveys of bands 1a - band 6b (band 5b was not available because of 
technical problems) and dedicated line observations of carbon monoxide
and its isotopes, and water and its isotopes.  
The telescope was used in a dual-beam switch mode with the source
placed alternatively in one of the two beams and cold sky in the other
beam, a method that yields very flat baselines
\citep{2010A&A...518L...6D,2010Roelfsema}.
A summary of the
observations is presented in Table~\ref{table1}.  
We note that Mars was not resolved, since its apparent diameter changed
from 8.1 to 8.3\arcsec{} during the observations. Thus, our observations
provide globally averaged quantities. 
The HCl multiplet at
1876 GHz and the \hiioii{} doublet at 1847 GHz were observed on
operational day (OD) 337 with 3746 and 2505 s integration times,
respectively, both in the upper sideband (USB) (see Table~\ref{table1}).
The \oii{} rotational transition at 774 GHz was observed twice on OD
334, once in the upper sideband with 2297 s and once in the lower
sideband (LSB) with 9289 s as a side product of a dedicated line
observation in the USB. The first set of data was available about a week
after the observations and was processed with the standard HIPE v3.0.1
modules \citep{2010HIPE} up to level 2. This data set remained incomplete
at the start of our study, for instance the data of the high resolution spectrometer (HRS)
was only partly available and pointing products therein had no
entries, thus, we analyzed only the wide band spectrometer (WBS) data.
This has no impact on the accuracy of the results presented in this
paper, although HRS data will be useful for future work including the
retrieval of vertical profiles.
Since the absolute flux calibration in the data set we obtained
from the \herschel{} Science Archive was still in progress, the
line-to-continuum ratio was analyzed rather than the absolute brightness
temperatures, as is standard for ground-based and other
\herschel{} observations \citep{2010A&A...518L.152L,2010A&A...518L.151S}.

\section{Analysis and discussion}

\begin{figure}
  \centering
  \includegraphics[trim = 0 0 93mm 0,clip,width=0.38\textwidth]{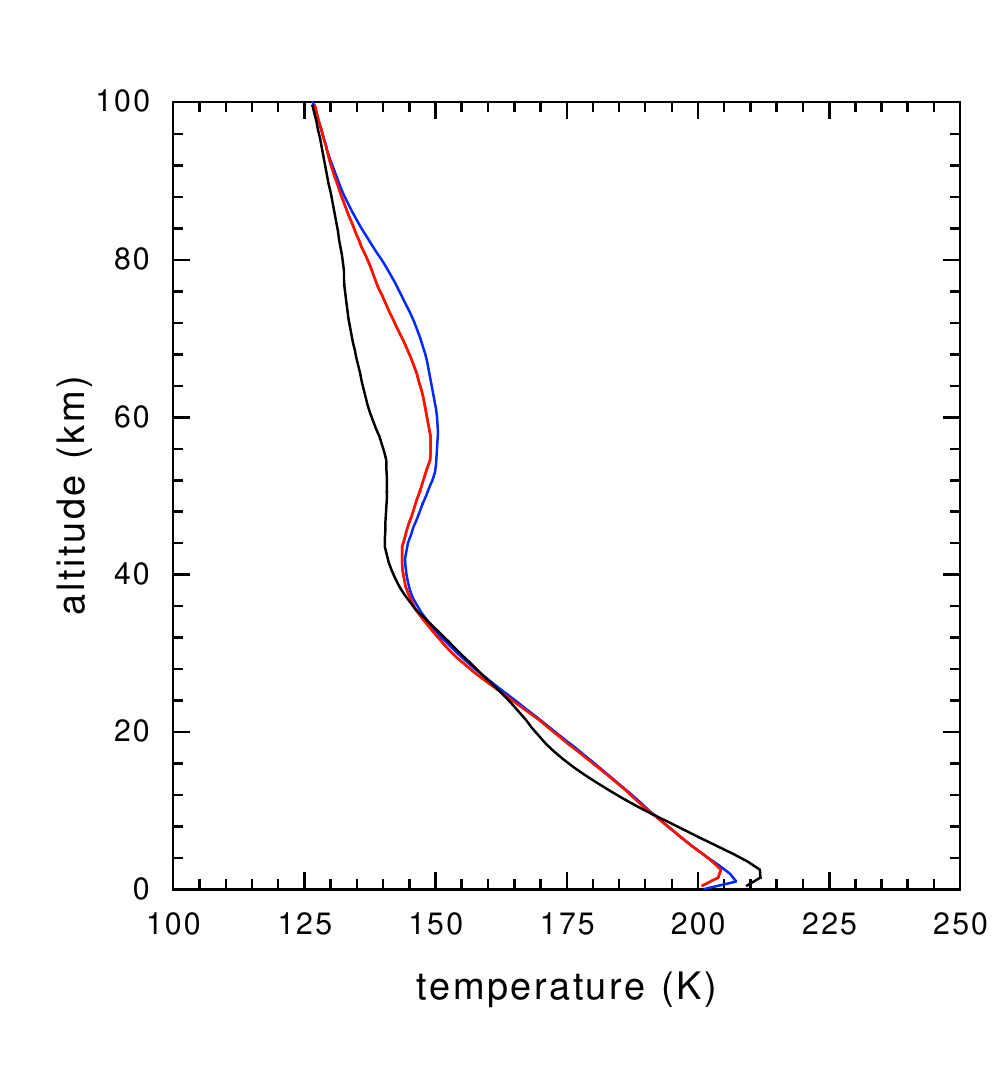}
  \caption{Temperature profiles predicted by EMCD (blue)
  \citep{1999JGR...10424155F,1999JGR...10424177L}, MAOAM (red)
  \citep{2005JGRE..11011008H,2007Icar..186...97M}, and retrieved vertical
  profile from simultaneous observations of \xiiico{} and \cxviiio{}.}
  \label{fig:temp}
\end{figure}

Compared to cometary observations of HIFI \citep{2010A&A...518L.150H,2010Wild},
the baseline ripple on the Mars observations is rather large, (as
frequently experienced by ground-based telescope observations of
planets), because of its strong continuum emission. While in the
cometary case the baseline ripple has been removed with a polynomial
fit, in the case of Mars we determined the baseline frequencies by a
normalized periodogram according to \citet{1976Ap&SS..39..447L} and
subtracted them from the original spectrum. This was applied
separately for horizontal and vertical polarization. After removal of
the baseline ripple, both polarizations were averaged.  In the case of
\oii{} observations, we found that the line strengths in both
sidebands were the same, and we therefore averaged the spectra obtained in
both sidebands. 

The observed spectral lines were modeled using a standard radiative
transfer code: Mars was assumed to be a perfect sphere surrounded by a set
of a hundred concentric atmospheric layers each of 1 km thickness
\citep[compare][]{2008P&SS...56.1368R}.  Within each layer, the
atmospheric temperature, pressure, and volume mixing ratio of carbon
monoxide were assumed to be constant.  The surface continuum emission was
modeled as black-body emission using a temperature distribution falling
off towards the edge of the apparent disk according to $T(\alpha) = T_0
\times (1- 0.2 \times (1-\cos(\alpha)))$, with $\alpha$ running from 0 -
90 ° across the apparent disk \citep[see also][]{2008A&A...489..795C}.
The disk-averaged emission was obtained
by integrating over the apparent disk using sixty four concentric rings
distributed unevenly over the disk and the limb region. The variation in
the path lengths through the atmosphere were fully taken into
account when calculating the radiation transfer of each ring. In our
model, the total continuum flux emitted by the surface depends purely on
the choice of the temperature $T_0$, which defines the temperature scale
for the temperature profile to be retrieved. We adjusted $T_0$ in
such a way to match exactly the total flux of about 4230 Jy predicted by
the `Mars continuum model' provided by \cite{2008Lellouch}. 

The absorption coefficients of the spectral lines were calculated using
the JPL spectral line catalog using the terrestrial isotopic ratios.
Pressure broadening coefficients for HCl and \hiioii{} were available only for
air, while they have been measured in the laboratory in a \coii{} atmosphere for
\oii{}. Most lab measurements display greater pressure broadening in a
\coii{} atmosphere. Its impact on the determination of upper limits is small. A
50\% increase in the pressure broadening coefficient leads to an
increase in the upper limit of 10 - 20\%.

For the retrieval of the mean volume mixing ratio of the three
molecules, we applied the temperature profile derived from HIFI
observations of \xiiico{} and \xiicxviiio{} during OD 334 \citep[][this
issue]{2010MarsCO} shown in Fig.~\ref{fig:temp}. 

\subsection{HCl}

\begin{figure}
  \centering
  \includegraphics[width=0.38\textwidth]{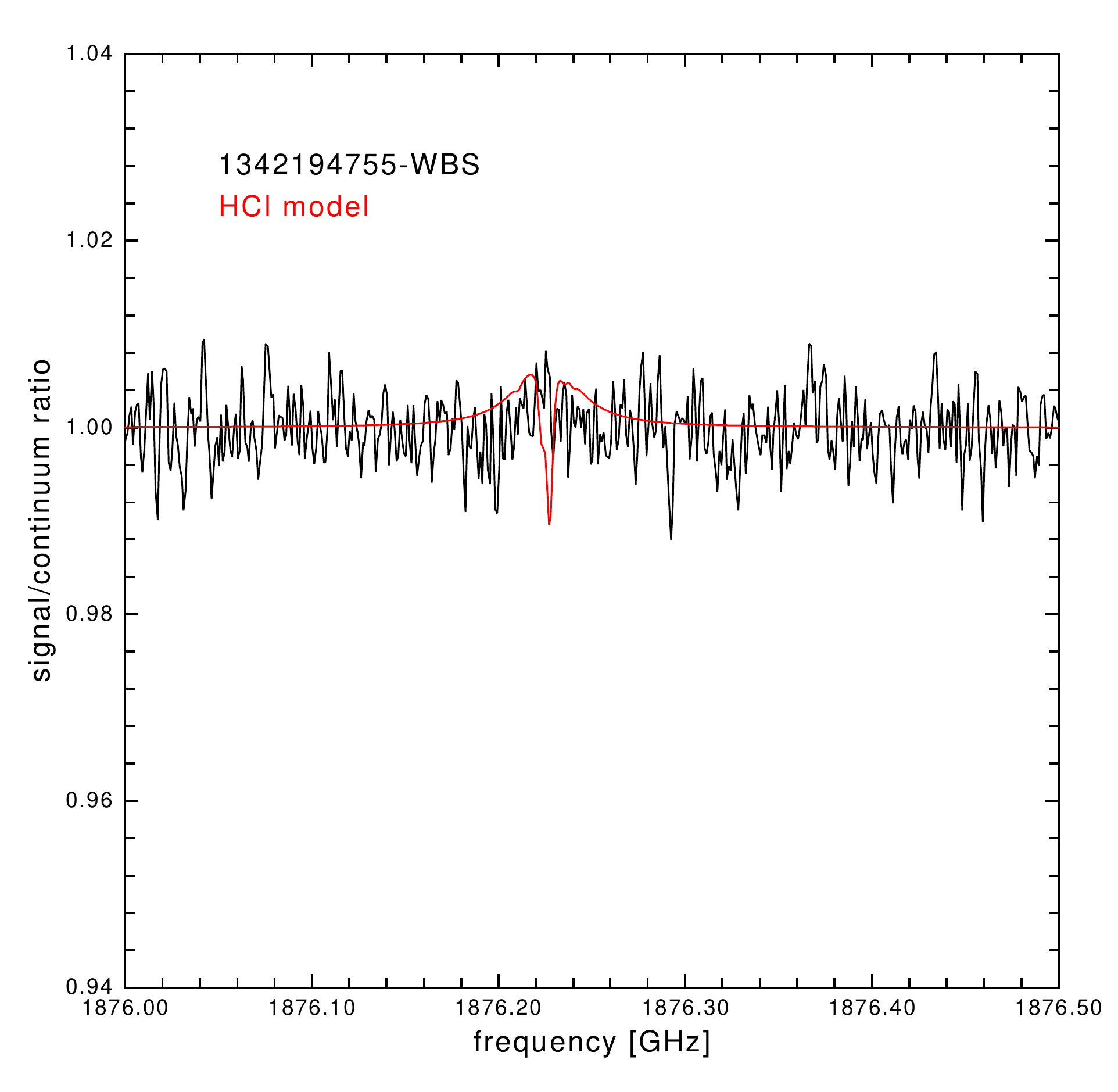}
  \caption{Observation of HCl centered around 1876 GHz and inserted
  model calculation (red) for a constant volume mixing ratio of 300
  ppt.}
  \label{fig:hcl}
\end{figure}

Figure~\ref{fig:hcl} shows the result of the 3746 s integration time on
the 1876 GHz H$^{35}$Cl line. We have inserted a modeled spectrum of HCl
assuming a constant volume mixing ratio of 300 ppt. HCl was obviously
not detected. If we define a line amplitude of $2\sigma$ as the upper
limit, we derive 200 ppt for HCl. This is one order of magnitude lower
than the upper limit derived by \citep{1997JGR...102.6525K} from IR
observations. We found no evidence of recent volcanic activity
or outgassing from a hot spot on Mars.  Nevertheless, the absence of HCl
does not preclude extant Martian volcanic activity.

\subsection{\hiioii{}}

\begin{figure}
  \centering
  \includegraphics[width=0.38\textwidth]{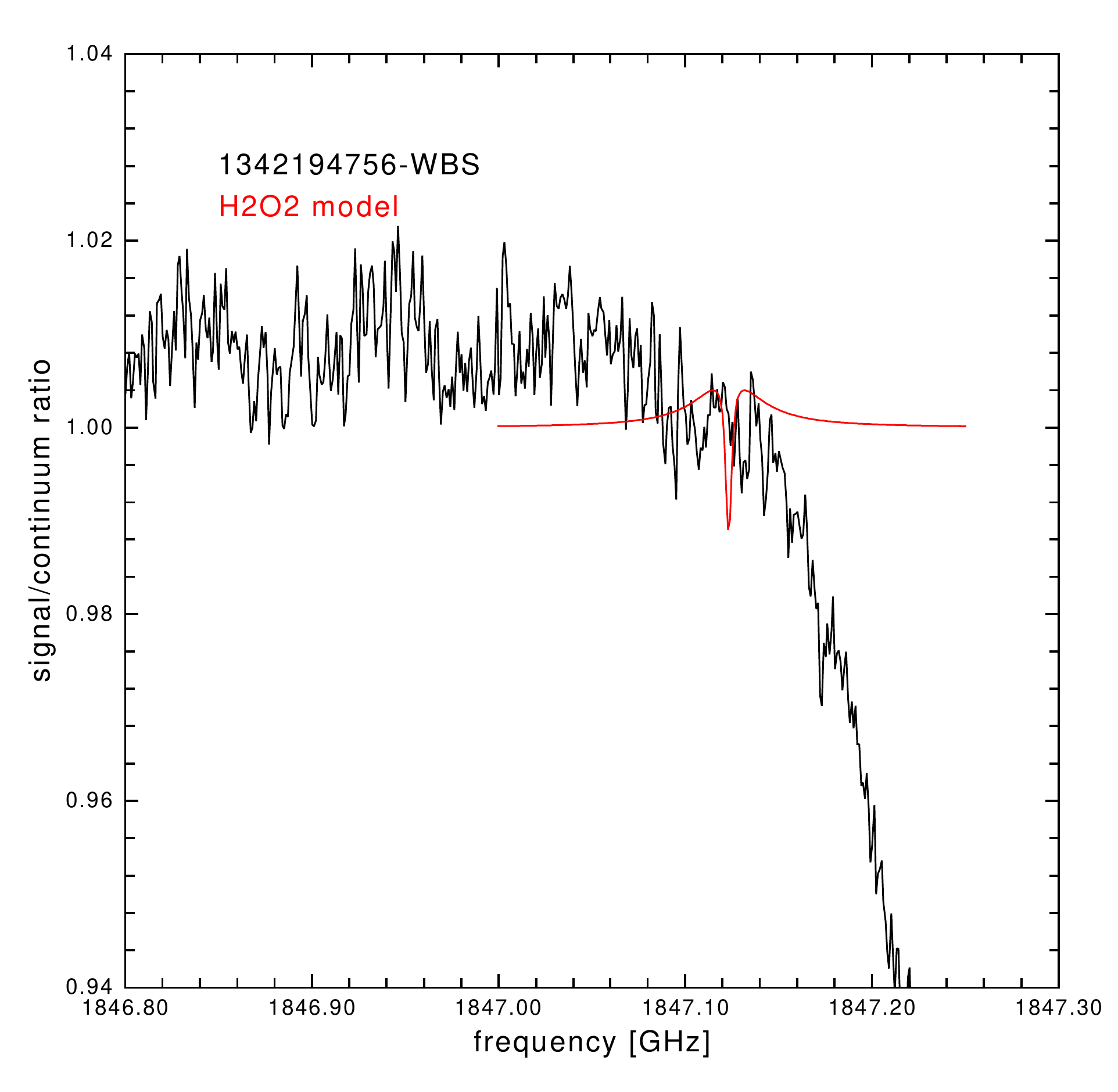}
  \caption{Observation of \hiioii{} spectrum at 1847 GHz  in the upper
  sideband and inserted model calculation (red) for a constant volume
  mixing ratio of 3 ppb. The strong absorption feature is CO (16-15) in
  the lower sideband.}
  \label{fig:h2o2}
\end{figure}

Figure~\ref{fig:h2o2} shows the result of the \hiioii{} observation on
1847 GHz in the upper sideband. The integration time was 2505 s. The
strong absorption feature is the CO (16-15) line. Since the line is in
the lower sideband centered around 1841 GHz, it does not absorb any
features of the \hiioii{} line. We did not detect any \hiioii{}. A
modeled \hiioii{} spectrum with a constant volume mixing ratio of 4 ppb
has been inserted into the measured spectrum. We deduced a 2-$\sigma$
upper limit of less than 3 ppb of \hiioii{}.  At first glance, this value
seems far too low taking into account former observations
providing 20--40 ppb (see Introduction). On the other hand \hiioii{}, is
connected to the water cycle and its high variability.
\citet{2009Icar..201..564K} compared the annual variability of \hiioii{}
based on observations and model calculations averaged over $\pm\
35\degr$ around the subsolar latitude.  Unfortunately, no other
observation for $L_s = 78\degr$ is available.  The model calculations
provided predictions for this season
\citep{2006Icar..185..153K,2009Icar..201..564K,2007Icar..188...18M,2008Natur.454..971L},
but all overestimated the volume mixing ratio compared to  our
observation.  \citet{2008Natur.454..971L} found about 10 ppb,
\citet{2007Icar..188...18M} for $L_s = 90\degr$ about 15 ppm and even
the lowest value of $\sim 5$ ppb calculated by
\citet{2009Icar..201..564K} is above the upper limit of our observation.
Nevertheless, the photochemical models predict lowest \hiioii{} values
for the season between $L_s = 70\degr$ and $80\degr$.  
Water vapour and its photolysis products are subject to solar cycle
variations \citep{2010JGRD..11500I17H}. A low Lyman-alpha flux (observations
were performed shortly after the solar minimum) may be consistent with less
than average production of \hiioii{} in the Martian atmosphere and
explain a negative deviation from the model values.

\subsection{\oii{}}

The upper panel of Fig. 4 shows the HIFI observation of the 774 GHz
\oii{} line -- the first submm detection of \oii{} in Mars --
and a model fit of a constant volume mixing
ratio. The best fit provides a volume mixing ratio of
$1400 \pm 120$ ppm.  This value fits within the error limits to the
value of 1300 ppm
derived in 1972.  We investigated the sensitivity of the
pressure broadening coefficient to this value. We initially applied  the
data from \citet{2003JMoSp.217..282G} for \oii{} in air: 1.62 MHz hPa$^{-1}$
(half width half maximum, HWHM). Taking into account the higher
molecular mass of \coii{} as the main collider compared with air, we
multiplied the pressure broadening coefficients in 0.1 hPa steps from
1.1 to 2 and found the best fit of the model to the observation for a
factor of 1.2, corresponding to 1.95 MHz hPa$^{-1}$ (HWHM). We note that the
mixing ratio was not found to be very sensitive to these changes, the
retrieved value always remaining within the error limits. The pressure
broadening factor of 1.2 is smaller than the factor of 1.4 (with \coii{}
rather than air being the main collider) for CO that has been found in
laboratory measurements \citep[e.g.][]{2009JQSRT.110..628D}.  The
quality of the observation is excellent, the signal-to-noise ratio 
being higher than 300. Unfortunately, the fit is not optimal. The model
underestimates the emission feature and overestimates the depth of the
absorption peak.  This indicates that the assumption of a constant
volume mixing ratio may not be correct. Deviations from the constant
profile seem to  be positive in the lower and negative in the upper
atmosphere. Future work will focus on the vertical profile of \oii{}.

\begin{figure}
   \centering
   \includegraphics[width=0.45\textwidth]{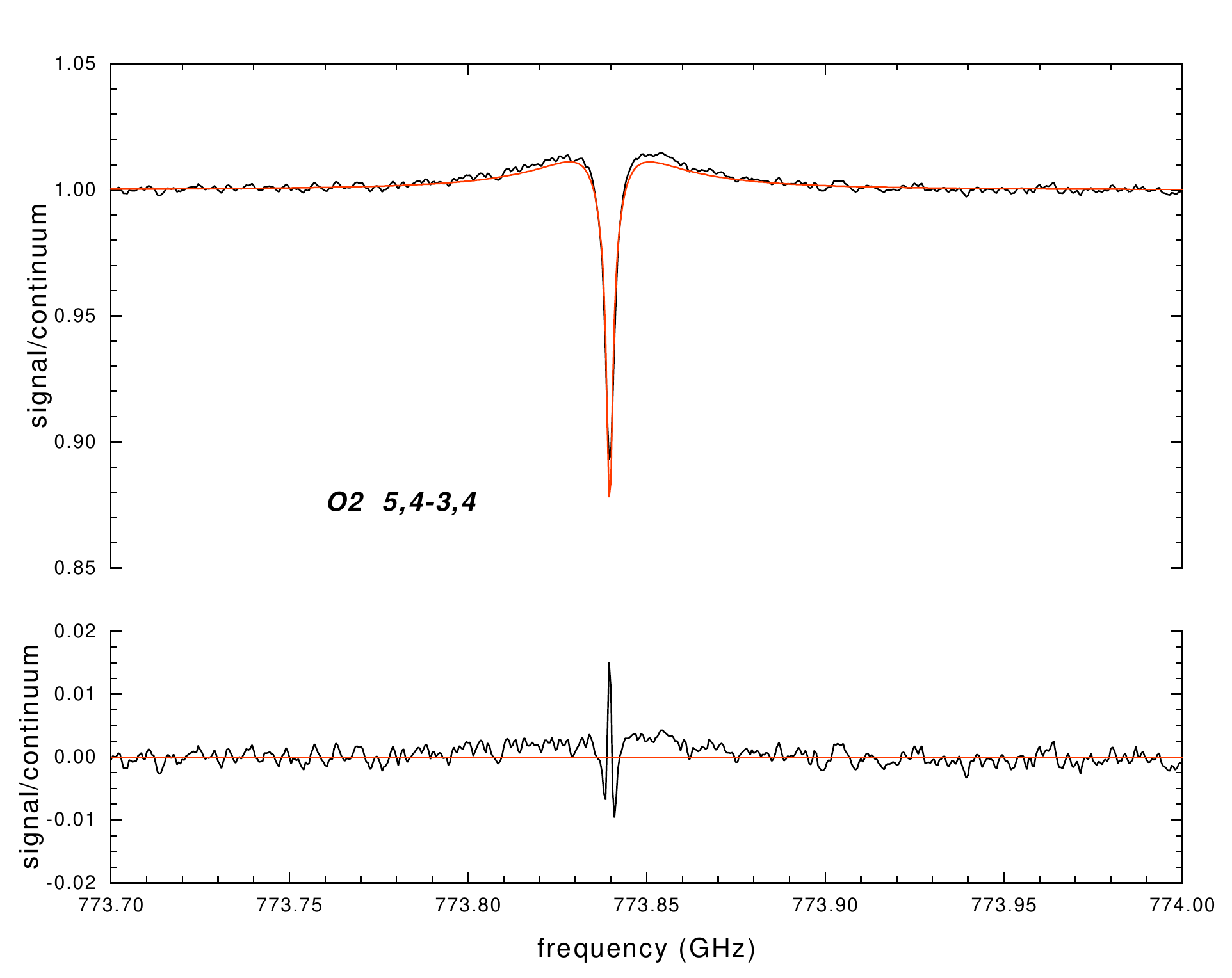}
   \caption{
   Observation of \oii{} at 774 GHz. The best fit of a constant altitude
   profile infers a volume mixing ratio of $1400 \pm 120$ ppm.
   The lower panel shows the difference between observation and model.
      }
\end{figure}

\section{Summary}

We have presented initial results for HIFI observations of the Martian
atmosphere on HCl, \hiioii{}, and \oii{}. The upper limit of 200 ppt
volume mixing ratio determined for HCl is one order of magnitude below
the previous value. There is no indication of present volcanic activity.
The upper limit to \hiioii{} of 2 ppb is remarkably low compared with
former detections. However, this observation is the first one around
$L_s = 77\degr$, a season where photochemical models predict the annual
minimum of \hiioii{}.  Future HIFI observations of \hiioii{} during
other solar longitudes will provide additional constraints on photochemical
models. The \oii{} volume mixing ratio of $1400 \pm 120$ ppm agrees with
former ground-based observations. The assumption of a constant vertical
profile does not lead to an optimal fit of the model to the
observations. The residuals suggest an oxygen  fall off  with height.
Future work will focus on the retrieval of the vertical \oii{} profile.

\begin{acknowledgements}
HIFI has been designed and built by a consortium of institutes and university
departments from across Europe, Canada and the United States under the
leadership of SRON Netherlands Institute for Space Research, Groningen, The
Netherlands and with major contributions from Germany, France and the US.
Consortium members are: Canada: CSA, U.Waterloo; France: CESR, LAB, LERMA,
IRAM; Germany: KOSMA, MPIfR, MPS; Ireland, NUI Maynooth; Italy: ASI, IFSI-INAF,
Osservatorio Astrofisico di Arcetri-INAF; Netherlands: SRON, TUD; Poland: CAMK,
CBK; Spain: Observatorio Astronómico Nacional (IGN), Centro de Astrobiología
(CSIC-INTA). Sweden: Chalmers University of Technology - MC2, RSS \& GARD;
Onsala Space Observatory; Swedish National Space Board, Stockholm University -
Stockholm Observatory; Switzerland: ETH Zurich, FHNW; USA: Caltech, JPL, NHSC.
HIPE is a joint development by the \herschel{} Science Ground Segment Consortium,
consisting of ESA, the NASA \herschel{} Science Center, and the HIFI, PACS and
SPIRE consortia.
This development has been supported by national funding agencies: CEA, CNES,
CNRS (France); ASI (Italy); DLR (Germany).
Additional funding support for some instrument activities has been provided by
ESA.
\end{acknowledgements}

\bibliographystyle{aa}
\bibliography{ads,preprints}

\begin{thebibliography}{45}
\expandafter\ifx\csname natexlab\endcsname\relax\def\natexlab#1{#1}\fi

\bibitem[{{Atreya} \& {Gu}(1994)}]{1994JGR....9913133A}
{Atreya}, S.~K. \& {Gu}, Z.~G. 1994, \jgr, 99, 13133

\bibitem[{{Atreya} {et~al.}(2006){Atreya}, {Wong}, {Renno}, {Farrell},
  {Delory}, {Sentman}, {Cummer}, {Marshall}, {Rafkin}, \&
  {Catling}}]{2006AsBio...6..439A}
{Atreya}, S.~K., {Wong}, A., {Renno}, N.~O., {et~al.} 2006, Astrobiology, 6,
  439

\bibitem[{{Barker}(1972)}]{1972Natur.238..447B}
{Barker}, E.~S. 1972, \nat, 238, 447

\bibitem[{{Belton} \& {Hunten}(1968)}]{1968ApJ...153..963B}
{Belton}, M.~J.~S. \& {Hunten}, D.~M. 1968, \apj, 153, 963

\bibitem[{{Bertaux} {et~al.}(2009){Bertaux}, {Vandaele}, {Korablev},
  {Montmessin}, {Marcq}, {Chaufray}, {Wilquet}, {Fedorova}, {Mahieux},
  {Belyaev}, {Villard}, {Gerard}, {Royer}, \&
  {Quemerais}}]{2009DPS....41.6002B}
{Bertaux}, J., {Vandaele}, A., {Korablev}, O., {et~al.} 2009, in AAS/DPS,
  Vol.~41, 60.02

\bibitem[{{Carleton} \& {Traub}(1972)}]{1972Sci...177..988C}
{Carleton}, N.~P. \& {Traub}, W.~A. 1972, Science, 177, 988

\bibitem[{{Cavali{\'e}} {et~al.}(2008){Cavali{\'e}}, {Billebaud}, {Encrenaz},
  {Dobrijevic}, {Brillet}, {Forget}, \& {Lellouch}}]{2008A&A...489..795C}
{Cavali{\'e}}, T., {Billebaud}, F., {Encrenaz}, T., {et~al.} 2008, \aap, 489,
  795

\bibitem[{{Clancy} {et~al.}(2004){Clancy}, {Sandor}, \&
  {Moriarty-Schieven}}]{2004Icar..168..116C}
{Clancy}, R.~T., {Sandor}, B.~J., \& {Moriarty-Schieven}, G.~H. 2004, Icarus,
  168, 116

\bibitem[{{Crewell} {et~al.}(1994){Crewell}, {K{\"u}nzi}, {Nett}, {Wehr}, \&
  {Hartogh}}]{1994GeoRL..21.1267C}
{Crewell}, S., {K{\"u}nzi}, K., {Nett}, H., {Wehr}, T., \& {Hartogh}, P. 1994,
  \grl, 21, 1267

\bibitem[{{de Graauw} {et~al.}(2010){de Graauw}, {Helmich}, {Phillips},
  {Stutzki}, {Caux}, {Whyborn}, {Dieleman}, {Roelfsema}, {Aarts}, {Assendorp},
  {Bachiller}, {Baechtold}, {Barcia}, {Beintema}, {Belitsky}, {Benz}, {Bieber},
  {Boogert}, {Borys}, {Bumble}, {Ca{\"i}s}, {Caris}, {Cerulli-Irelli},
  {Chattopadhyay}, {Cherednichenko}, {Ciechanowicz}, {Coeur-Joly}, {Comito},
  {Cros}, {de Jonge}, {de Lange}, {Delforges}, {Delorme}, {den Boggende},
  {Desbat}, {Diez-Gonz{\'a}lez}, {di Giorgio}, {Dubbeldam}, {Edwards},
  {Eggens}, {Erickson}, {Evers}, {Fich}, {Finn}, {Franke}, {Gaier}, {Gal},
  {Gao}, {Gallego}, {Gauffre}, {Gill}, {Glenz}, {Golstein}, {Goulooze},
  {Gunsing}, {G{\"u}sten}, {Hartogh}, {Hatch}, {Higgins}, {Honingh}, {Huisman},
  {Jackson}, {Jacobs}, {Jacobs}, {Jarchow}, {Javadi}, {Jellema}, {Justen},
  {Karpov}, {Kasemann}, {Kawamura}, {Keizer}, {Kester}, {Klapwijk}, {Klein},
  {Kollberg}, {Kooi}, {Kooiman}, {Kopf}, {Krause}, {Krieg}, {Kramer},
  {Kruizenga}, {Kuhn}, {Laauwen}, {Lai}, {Larsson}, {Leduc}, {Leinz}, {Lin},
  {Liseau}, {Liu}, {Loose}, {L{\'o}pez-Fernandez}, {Lord}, {Luinge}, {Marston},
  {Mart{\'{\i}}n-Pintado}, {Maestrini}, {Maiwald}, {McCoey}, {Mehdi}, {Megej},
  {Melchior}, {Meinsma}, {Merkel}, {Michalska}, {Monstein}, {Moratschke},
  {Morris}, {Muller}, {Murphy}, {Naber}, {Natale}, {Nowosielski}, {Nuzzolo},
  {Olberg}, {Olbrich}, {Orfei}, {Orleanski}, {Ossenkopf}, {Peacock}, {Pearson},
  {Peron}, {Phillip-May}, {Piazzo}, {Planesas}, {Rataj}, {Ravera}, {Risacher},
  {Salez}, {Samoska}, {Saraceno}, {Schieder}, {Schlecht}, {Schl{\"o}der},
  {Schm{\"u}lling}, {Schultz}, {Schuster}, {Siebertz}, {Smit}, {Szczerba},
  {Shipman}, {Steinmetz}, {Stern}, {Stokroos}, {Teipen}, {Teyssier}, {Tils},
  {Trappe}, {van Baaren}, {van Leeuwen}, {van de Stadt}, {Visser}, {Wildeman},
  {Wafelbakker}, {Ward}, {Wesselius}, {Wild}, {Wulff}, {Wunsch}, {Tielens},
  {Zaal}, {Zirath}, {Zmuidzinas}, \& {Zwart}}]{2010A&A...518L...6D}
{de Graauw}, T., {Helmich}, F.~P., {Phillips}, T.~G., {et~al.} 2010, \aap, 518,
  L6

\bibitem[{{de Val-Borro} {et~al.}(2010){de Val-Borro}, {Hartogh}, {Crovisier},
  {Bockel{\'e}e-Morvan}, {Biver}, {Lis}, {Moreno}, \& {Jarchow}}]{2010Wild}
{de Val-Borro}, M., {Hartogh}, P., {Crovisier}, J., {et~al.} 2010, \aap\ in
  press

\bibitem[{{Dick} {et~al.}(2009){Dick}, {Drouin}, {Crawford}, \&
  {Pearson}}]{2009JQSRT.110..628D}
{Dick}, M.~J., {Drouin}, B.~J., {Crawford}, T.~J., \& {Pearson}, J.~C. 2009,
  Journal of Quantitative Spectroscopy and Radiative Transfer, 110, 628

\bibitem[{{Encrenaz} {et~al.}(2004){Encrenaz}, {B{\'e}zard}, {Greathouse},
  {Richter}, {Lacy}, {Atreya}, {Wong}, {Lebonnois}, {Lef{\`e}vre}, \&
  {Forget}}]{2004Icar..170..424E}
{Encrenaz}, T., {B{\'e}zard}, B., {Greathouse}, T.~K., {et~al.} 2004, Icarus,
  170, 424

\bibitem[{England \& Hrubes(2001)}]{2001England}
England, C. \& Hrubes, J.~D. 2001, MARRS-Mars Atmosphere Resource Recovery
  System, study report available at \url{http://www.niac.usra.edu}

\bibitem[{{England} \& {Hrubes}(2004)}]{2004otp..work.3009E}
{England}, C. \& {Hrubes}, J.~D. 2004, in Workshop on Oxygen in the Terrestrial
  Planets, ed. {J.~Jones \& C.~Herd}, 3009

\bibitem[{{Forget} {et~al.}(1999){Forget}, {Hourdin}, {Fournier}, {Hourdin},
  {Talagrand}, {Collins}, {Lewis}, {Read}, \& {Huot}}]{1999JGR...10424155F}
{Forget}, F., {Hourdin}, F., {Fournier}, R., {et~al.} 1999, \jgr, 104, 24155

\bibitem[{{Golubiatnikov} \& {Krupnov}(2003)}]{2003JMoSp.217..282G}
{Golubiatnikov}, G.~Y. \& {Krupnov}, A.~F. 2003, Journal of Molecular
  Spectroscopy, 217, 282

\bibitem[{{Hartogh} {et~al.}(2010{\natexlab{a}}){Hartogh}, {B\l{}ecka},
  {Jarchow}, {Sagawa}, {Lellouch}, {de Val-Borro}, \& {Rengel}}]{2010MarsCO}
{Hartogh}, P., {B\l{}ecka}, M.~I., {Jarchow}, C., {et~al.} 2010{\natexlab{a}},
  \aap\ this issue

\bibitem[{{Hartogh} {et~al.}(2010{\natexlab{b}}){Hartogh}, {Crovisier}, {de
  Val-Borro}, {Bockel{\'e}e-Morvan}, {Biver}, {Lis}, {Moreno}, {Jarchow},
  {Rengel}, {Emprechtinger}, {Szutowicz}, {Banaszkiewicz}, {Bensch}, {Blecka},
  {Cavali{\'e}}, {Encrenaz}, {Jehin}, {K{\"u}ppers}, {Lara}, {Lellouch},
  {Swinyard}, {Vandenbussche}, {Bergin}, {Blake}, {Blommaert}, {Cernicharo},
  {Decin}, {Encrenaz}, {de Graauw}, {Hutsemekers}, {Kidger}, {Manfroid},
  {Medvedev}, {Naylor}, {Schieder}, {Thomas}, {Waelkens}, {Roelfsema},
  {Dieleman}, {G{\"u}sten}, {Klein}, {Kasemann}, {Caris}, {Olberg}, \&
  {Benz}}]{2010A&A...518L.150H}
{Hartogh}, P., {Crovisier}, J., {de Val-Borro}, M., {et~al.}
  2010{\natexlab{b}}, \aap, 518, L150

\bibitem[{{Hartogh} {et~al.}(2009){Hartogh}, {Lellouch}, {Crovisier},
  {Banaszkiewicz}, {Bensch}, {Bergin}, {Billebaud}, {Biver}, {Blake}, {Blecka},
  {Blommaert}, {Bockel{\'e}e-Morvan}, {Cavali{\'e}}, {Cernicharo}, {Courtin},
  {Davis}, {Decin}, {Encrenaz}, {Encrenaz}, {Gonz{\'a}lez}, {de Graauw},
  {Hutsem{\'e}kers}, {Jarchow}, {Jehin}, {Kidger}, {K{\"u}ppers}, {de Lange},
  {Lara}, {Lis}, {Lorente}, {Manfroid}, {Medvedev}, {Moreno}, {Naylor},
  {Orton}, {Portyankina}, {Rengel}, {Sagawa}, {S{\'a}nchez-Portal}, {Schieder},
  {Sidher}, {Stam}, {Swinyard}, {Szutowicz}, {Thomas}, {Thornhill},
  {Vandenbussche}, {Verdugo}, {Waelkens}, \& {Walker}}]{2009P&SS...57.1596H}
{Hartogh}, P., {Lellouch}, E., {Crovisier}, J., {et~al.} 2009, \planss, 57,
  1596

\bibitem[{{Hartogh} {et~al.}(2005){Hartogh}, {Medvedev}, {Kuroda}, {Saito},
  {Villanueva}, {Feofilov}, {Kutepov}, \& {Berger}}]{2005JGRE..11011008H}
{Hartogh}, P., {Medvedev}, A.~S., {Kuroda}, T., {et~al.} 2005, JGR, 110, 11008

\bibitem[{{Hartogh} {et~al.}(2010{\natexlab{c}}){Hartogh}, {Sonnemann},
  {Grygalashvyly}, {Song}, {Berger}, \& {L{\"u}bken}}]{2010JGRD..11500I17H}
{Hartogh}, P., {Sonnemann}, G.~R., {Grygalashvyly}, M., {et~al.}
  2010{\natexlab{c}}, JGR, 115, D00117

\bibitem[{{Iwagami} {et~al.}(2008){Iwagami}, {Ohtsuki}, {Tokuda}, {Ohira},
  {Kasaba}, {Imamura}, {Sagawa}, {Hashimoto}, {Takeuchi}, {Ueno}, \&
  {Okumura}}]{2008P&SS...56.1424I}
{Iwagami}, N., {Ohtsuki}, S., {Tokuda}, K., {et~al.} 2008, \planss, 56, 1424

\bibitem[{{Kaplan} {et~al.}(1969){Kaplan}, {Connes}, \&
  {Connes}}]{1969ApJ...157L.187K}
{Kaplan}, L.~D., {Connes}, J., \& {Connes}, P. 1969, \apjl, 157, L187+

\bibitem[{{Krasnopolsky}(1993)}]{1993Icar..101..313K}
{Krasnopolsky}, V.~A. 1993, Icarus, 101, 313

\bibitem[{{Krasnopolsky}(2006)}]{2006Icar..185..153K}
{Krasnopolsky}, V.~A. 2006, Icarus, 185, 153

\bibitem[{{Krasnopolsky}(2009)}]{2009Icar..201..564K}
{Krasnopolsky}, V.~A. 2009, Icarus, 201, 564

\bibitem[{{Krasnopolsky} {et~al.}(1997){Krasnopolsky}, {Bjoraker}, {Mumma}, \&
  {Jennings}}]{1997JGR...102.6525K}
{Krasnopolsky}, V.~A., {Bjoraker}, G.~L., {Mumma}, M.~J., \& {Jennings}, D.~E.
  1997, \jgr, 102, 6525

\bibitem[{{Lef{\`e}vre} {et~al.}(2008){Lef{\`e}vre}, {Bertaux}, {Clancy},
  {Encrenaz}, {Fast}, {Forget}, {Lebonnois}, {Montmessin}, \&
  {Perrier}}]{2008Natur.454..971L}
{Lef{\`e}vre}, F., {Bertaux}, J., {Clancy}, R.~T., {et~al.} 2008, \nat, 454,
  971

\bibitem[{Lellouch \& Amri(2008)}]{2008Lellouch}
Lellouch, E. \& Amri, H. 2008,
  \url{http://www.lesia.obspm.fr/perso/emmanuel-lellouch/mars/}

\bibitem[{{Lellouch} {et~al.}(2010){Lellouch}, {Hartogh}, {Feuchtgruber},
  {Vandenbussche}, {de Graauw}, {Moreno}, {Jarchow}, {Cavali{\'e}}, {Orton},
  {Banaszkiewicz}, {Blecka}, {Bockel{\'e}e-Morvan}, {Crovisier}, {Encrenaz},
  {Fulton}, {K{\"u}ppers}, {Lara}, {Lis}, {Medvedev}, {Rengel}, {Sagawa},
  {Swinyard}, {Szutowicz}, {Bensch}, {Bergin}, {Billebaud}, {Biver}, {Blake},
  {Blommaert}, {Cernicharo}, {Courtin}, {Davis}, {Decin}, {Encrenaz},
  {Gonzalez}, {Jehin}, {Kidger}, {Naylor}, {Portyankina}, {Schieder}, {Sidher},
  {Thomas}, {de Val-Borro}, {Verdugo}, {Waelkens}, {Walker}, {Aarts}, {Comito},
  {Kawamura}, {Maestrini}, {Peacocke}, {Teipen}, {Tils}, \&
  {Wildeman}}]{2010A&A...518L.152L}
{Lellouch}, E., {Hartogh}, P., {Feuchtgruber}, H., {et~al.} 2010, \aap, 518,
  L152

\bibitem[{{Lewis} {et~al.}(1999){Lewis}, {Collins}, {Read}, {Forget},
  {Hourdin}, {Fournier}, {Hourdin}, {Talagrand}, \&
  {Huot}}]{1999JGR...10424177L}
{Lewis}, S.~R., {Collins}, M., {Read}, P.~L., {et~al.} 1999, \jgr, 104, 24177

\bibitem[{{Lomb}(1976)}]{1976Ap&SS..39..447L}
{Lomb}, N.~R. 1976, \apss, 39, 447

\bibitem[{{Medvedev} \& {Hartogh}(2007)}]{2007Icar..186...97M}
{Medvedev}, A.~S. \& {Hartogh}, P. 2007, Icarus, 186, 97

\bibitem[{{Moudden} \& {McConnell}(2007)}]{2007Icar..188...18M}
{Moudden}, Y. \& {McConnell}, J.~C. 2007, Icarus, 188, 18

\bibitem[{{Mumma} {et~al.}(2009){Mumma}, {Villanueva}, {Novak}, {Hewagama},
  {Bonev}, {DiSanti}, {Mandell}, \& {Smith}}]{2009Sci...323.1041M}
{Mumma}, M.~J., {Villanueva}, G.~L., {Novak}, R.~E., {et~al.} 2009, Science,
  323, 1041

\bibitem[{{Nair} {et~al.}(1994){Nair}, {Allen}, {Anbar}, {Yung}, \&
  {Clancy}}]{1994Icar..111..124N}
{Nair}, H., {Allen}, M., {Anbar}, A.~D., {Yung}, Y.~L., \& {Clancy}, R.~T.
  1994, Icarus, 111, 124

\bibitem[{Ott(2010)}]{2010HIPE}
Ott, S. 2010, ASP Conference Series, Astronomical Data Analysis Software and
  Systems XIX, Y. Mizumoto, K.-I. Morita, and M. Ohishi, eds., in press

\bibitem[{{Owen} {et~al.}(1977){Owen}, {Biemann}, {Biller}, {Lafleur},
  {Rushneck}, \& {Howarth}}]{1977JGR....82.4635O}
{Owen}, T., {Biemann}, K., {Biller}, J.~E., {et~al.} 1977, \jgr, 82, 4635

\bibitem[{{Rengel} {et~al.}(2008){Rengel}, {Hartogh}, \&
  {Jarchow}}]{2008P&SS...56.1368R}
{Rengel}, M., {Hartogh}, P., \& {Jarchow}, C. 2008, \planss, 56, 1368

\bibitem[{{Roelfsema} {et~al.}(2010){Roelfsema}, {Helmich}, {Teyssier}, \& {et
  al.}}]{2010Roelfsema}
{Roelfsema}, P., {Helmich}, F., {Teyssier}, D., \& {et al.} 2010, \aap\ this
  issue

\bibitem[{{Swinyard} {et~al.}(2010){Swinyard}, {Hartogh}, {Sidher}, {Fulton},
  {Lellouch}, {Jarchow}, {Griffin}, {Moreno}, {Sagawa}, {Portyankina},
  {Blecka}, {Banaszkiewicz}, {Bockelee-Morvan}, {Crovisier}, {Encrenaz},
  {Kueppers}, {Lara}, {Lis}, {Medvedev}, {Rengel}, {Szutowicz},
  {Vandenbussche}, {Bensch}, {Bergin}, {Billebaud}, {Biver}, {Blake},
  {Blommaert}, {de Val-Borro}, {Cernicharo}, {Cavalie}, {Courtin}, {Davis},
  {Decin}, {Encrenaz}, {de Graauw}, {Jehin}, {Kidger}, {Leeks}, {Orton},
  {Naylor}, {Schieder}, {Stam}, {Thomas}, {Verdugo}, {Waelkens}, \&
  {Walker}}]{2010A&A...518L.151S}
{Swinyard}, B.~M., {Hartogh}, P., {Sidher}, S., {et~al.} 2010, \aap, 518, L151

\bibitem[{{Very}(1909)}]{1909Sci....30..678V}
{Very}, F.~W. 1909, Science, 30, 678

\bibitem[{{Wehr} {et~al.}(1995){Wehr}, {Crewell}, {K{\"u}nzi}, {Langen},
  {Nett}, {Urban}, \& {Hartogh}}]{1995JGR...10020957W}
{Wehr}, T., {Crewell}, S., {K{\"u}nzi}, K., {et~al.} 1995, \jgr, 100, 20957

\bibitem[{{Wong} {et~al.}(2003){Wong}, {Atreya}, \&
  {Encrenaz}}]{2003JGRE..108.5026W}
{Wong}, A., {Atreya}, S.~K., \& {Encrenaz}, T. 2003, Journal of Geophysical
  Research (Planets), 108, 5026

\end{thebibliography}

\end{document}